\documentstyle[12pt]{article}
\renewcommand{\theequation}{\thesection.\arabic{equation}}
\renewcommand{\thesubsection}{\thesection.\arabic{subsection}}

\newlength{\extraspace}
\newlength{\extraspaces}
\newcounter{dummy}

\newcommand{\be}{\begin{equation}
\addtolength{\abovedisplayskip}{\extraspaces}
\addtolength{\belowdisplayskip}{\extraspaces}
\addtolength{\abovedisplayshortskip}{\extraspace}
\addtolength{\belowdisplayshortskip}{\extraspace}}
\newcommand{\ee}{\end{equation}}

\newcommand{\ba}{\begin{eqnarray}
\addtolength{\abovedisplayskip}{\extraspaces}
\addtolength{\belowdisplayskip}{\extraspaces}
\addtolength{\abovedisplayshortskip}{\extraspace}
\addtolength{\belowdisplayshortskip}{\extraspace}}
\newcommand{\ea}{\end{eqnarray}}

\newcommand{\baa}{
\addtocounter{equation}{1}
\setcounter{dummy}{\value{equation}}
\setcounter{equation}{0}
\renewcommand{\theequation}{\thesection.\arabic{dummy}\alph{equation}}
\begin{eqnarray}
\addtolength{\abovedisplayskip}{\extraspaces}
\addtolength{\belowdisplayskip}{\extraspaces}
\addtolength{\abovedisplayshortskip}{\extraspace}
\addtolength{\belowdisplayshortskip}{\extraspace}}
\newcommand{\eaa}{
\end{eqnarray}
\setcounter{equation}{\value{dummy}}
\renewcommand{\theequation}{\thesection.\arabic{equation}}}

\newcommand{\ban}{\begin{eqnarray*}
\addtolength{\abovedisplayskip}{\extraspaces}
\addtolength{\belowdisplayskip}{\extraspaces}
\addtolength{\abovedisplayshortskip}{\extraspace}
\addtolength{\belowdisplayshortskip}{\extraspace}}
\newcommand{\ean}{\end{eqnarray*}}

\newcommand{\newsection}[1]{
\vspace{13mm}
\pagebreak[3]
\addtocounter{section}{1}
\setcounter{equation}{0}
\setcounter{subsection}{0}
\setcounter{footnote}{0}
\begin{flushleft}
{\large\bf \thesection. #1}
\end{flushleft}
\nopagebreak
\smallskip
\nopagebreak}

\newcommand{\newsubsection}[1]{
\vspace{13mm}
\pagebreak[3]

\addtocounter{subsection}{1}
\addcontentsline{toc}{subsection}{\protect
\numberline{\arabic{section}.\arabic{subsection}}{#1}}
\noindent{\bf \thesubsection. #1}
\nopagebreak
\vspace{2mm}
\nopagebreak}

\newcommand{\ie}{{\it i.e.}}

\newcommand{\hf}{{\textstyle{1\over 2}}}

\newcommand{\is}{\! & \! = \! & \!}

\newcommand{\nonu}{\nonumber \\[1.5mm]}

\newcommand{\ra}{\rightarrow}

\newcommand{\figuur}[2]{
\begin{figure}
\vspace{#1mm}
\begin{center}
\setlength{\unitlength}{.9mm}
\raisebox{-20\unitlength}
{\mbox{\begin{picture}(70,65)(-30,-35)
\thicklines
\put(-74,5){\line(1,1){15}}
\put(-59,20){\line(1,0){30}}
\put(-59,20){\line(1,-1){30}}
\put(-44,-25){\line(-1,1){30}}
\put(-29,-10){\vector(-1,1){10}}
\thinlines
\put(-44,5){\vector(1,1){7.5}}
\put(-59,-10){\vector(1,1){7.5}}
\put(-29,20){\line(1,-1){15}}
\put(-29,20){\line(-1,-1){30}}
\put(-14,5){\line(-1,-1){30}}
\put(-59.5,19.5){\line(1,0){31}}
\put(-38,-4){\makebox(0,0){\small $p_+$}}
\put(-31,11){\makebox(0,0)
{\scriptsize{$x^{{}^-}\!\!\!\!=\!-{p_+\over\lambda^2}$}}}
\put(-7,5){\makebox(0,0){\scriptsize{$x^{{}^\pm} \!\!\!=\!\! \pm \infty$ }}}
\put(-70,15){\makebox(0,0){\scriptsize{$x^{{}^-} \!\!\!=\! 0$ }}}
\put(-67,-7){\makebox(0,0){\scriptsize{$x^{{}^+} \!\!\! =\!  0$ }}}
\put(-26,-13){\makebox(0,0){\footnotesize ${\cal I}_R^-$}}
\put(-19,15){\makebox(0,0){\footnotesize ${\cal I}_R^+$}}
\setlength{\unitlength}{.95mm}
\thicklines
\put(19,0){\line(1,1){15}}
\put(59,-10){\line(-1,1){35}}
\put(24,25){\line(1,0){20}}
\put(44,-25){\line(-1,1){25}}
\put(59,-10){\vector(-1,1){10}}
\thinlines
\put(69,0){\line(-1,1){25}}
\put(29,-10){\line(1,1){15}}
\put(29,-10){\vector(1,1){7.5}}
\put(34,+15){\line(1,1){10}}
\put(34,+15){\vector(1,1){5}}
\put(69,0){\line(-1,-1){25}}
\put(24.5,24.5){\line(1,0){20}}
\put(51,-5){\makebox(0,0){\small $p_+$}}
\put(18.5,25){\makebox(0,0)
{\scriptsize{$x^{{}^-}\!\!\!\! =\!{p_+\over\lambda^2}$}}}
\put(76,0){\makebox(0,0){\scriptsize{$x^{{}^\pm} \!\!\!=\!\! \pm \infty$ }}}
\put(24,10){\makebox(0,0){\scriptsize{$x^{{}^-} \!\!\!\!=\! 0$ }}}
\put(28,-13){\makebox(0,0){\scriptsize{$x^{{}^+} \!\!\! =\!  0$ }}}
\put(60,-16){\makebox(0,0){\footnotesize ${\cal I}_R^-$}}
\put(60,16){\makebox(0,0){\footnotesize ${\cal I}_R^+$}}
\end{picture}}}
\parbox{15cm}{\small #2}
\end{center}
\end{figure}}

\newcommand{\half}{{\textstyle{1\over 2}}}

\newcommand{\X}{{\mbox{\footnotesize $X$}}}
\newcommand{\PP}{{\mbox{\footnotesize $P$}}}
\newcommand{\Xt}{\tilde{\mbox{\footnotesize $X$}}}

\newcommand{\vac}{|0\rangle}
\newcommand{\Xh}{\hat{\X}}
\newcommand{\del}{\partial}

\begin{document}

\input epsf

\addtolength{\baselineskip}{.65mm}
\thispagestyle{empty}

\begin{flushright}
{\sc PUPT}-1395\\
{\sc IASSNS-HEP}-93/25\\
April 1993
\end{flushright}

\begin{center}
{\Large{Quantum Black Hole Evaporation.}}\\[10mm]
{\sc Kareljan Schoutens, Herman Verlinde}\\[3mm]
{\it Joseph Henry Laboratories\\[2mm]
 Princeton University, Princeton, NJ 08544}
\\[.4cm]
{ and}\\[.4cm]
{\sc Erik Verlinde}\\[3mm]
{\it School of Natural Sciences\\[2mm]
Institute for Advanced Study, Princeton, NJ 08540}
\\[10mm]

{\sc Abstract}
\end{center}
We investigate a recently proposed model for a full quantum description of
two-dimensi\-onal black hole evaporation, in which a reflecting boundary
condition is imposed in the strong coupling region.  It is shown that in this
model each initial state is mapped to a well-defined asymptotic out-state,
provided one performs a certain projection in the gravitational
zero mode sector. We find that for an incoming localized energy pulse,
the corresponding out-going state contains approximately thermal
radiation, in accordance with semi-classical predictions. In addition, our
model allows for certain acausal strong coupling effects near the singularity,
that give rise to corrections to the Hawking spectrum and restore the
coherence of the out-state. To an asymptotic observer these corrections
appear to originate from behind the receding apparent horizon and start
to influence the out-going state long before the black hole has emitted most
of its mass. Finally, by putting the system in a finite box,
we are able to derive some algebraic properties of the
scattering matrix and prove that the final state contains all initial
information.

\noindent

\vfill

\noindent

\vfill

\newpage
\newsection{Introduction}

Classically, black holes are absolutely black. They do not allow
any signals to leave their surface and forever hide themselves behind
a horizon without any means of communication with the outside world.
In quantum mechanics, however, this classical notion of a black hole needs
a fundamental revision. Hawking's discovery that quantum black holes can
evaporate by emitting thermal radiation \cite{hawking} shows that they are
neither absolutely black nor can hide themselves forever. Nevertheless,
the very mechanism by which black holes can lose their mass does
not seem to allow for any information to be carried out with it.
However drastic, it seems at present an almost foregone conclusion
that black hole evaporation leads to information loss
and to the evolution of pure states into mixed states.
Before accepting such a far reaching conclusion, however, it seems only
appropriate to make sure that other, more conservative, logical
possibilities are ruled out.

A simple toy model for studying some of these issues
is two-dimensional dilaton gravity coupled to $N$ massless scalar fields
\cite{cghs}.
This model exhibits an instability against gravitational collapse very similar
to the Einstein theory, but at the same time it is simple enough to allow an
explicit analysis of its quantum properties.  In this paper we will
further investigate a recently proposed method for quantizing dilaton
gravity with $N=24$ matter fields \cite{us2}.
Our goal is to use this model
to obtain a full quantum description of two-dimensional black hole
evaporation and to investigate whether the problem of information loss
can be avoided while retaining the correspondence with semi-classical physics.
This last requirement means concretely that for physically reasonable
initial states, a significant part of the out-going state should
describe almost thermal radiation, as this would be a true
signature that a black hole was actually formed.

This paper is organized as follows.
In section 2 we use a simple gravitational shockwave
picture to draw a parallel between dilaton gravity
and 't Hooft's model for a 3+1-dimensional black hole $S$-matrix.
The  quantization of $N=24$ dilaton gravity is reviewed in section 3. We
describe the physical spectrum and  formulate the boundary condition
that leads to the definition of the $S$-matrix. In section 4 we summarize
the interactions between the in- and out-going modes by means of a
certain `exchange algebra'. We use this algebra to analyze
some physical properties of the $S$-matrix and to make contact with
the semi-classical theory and to discuss the corrections to it.
In section 5 we use the analogy with
open string theory to put the model in a finite box. This allows us to
prove that the $S$-matrix indeed maps every in-state to a pure out-state.
In the last section we address some questions concerning the unitarity and
physical interpretation of the scattering matrix.

\newsection{The Black Hole Problem Miniaturized}

In this section we will use the two-dimensional dilaton gravity model
to illustrate some aspects of the black hole information paradox.
In particular we will summarize 't Hooft's argument \cite{thooft}
for the form of the
black hole scattering matrix in this simplified context
(see also \cite{peetal}). The purpose of this discussion is to
provide a physical context for some of the calculations in the
coming sections, and to motivate the need for and the form of the boundary
conditions.
Moreover, it turns out that this simplified version of 't Hooft's
$S$-matrix appears as an overall factor
in the full quantum scattering matrix of dilaton gravity, and
therefore similar issues arise when one tries to find their
proper physical interpretation.
The reasonings in this section
are semi-classical and somewhat intuitive, as the full quantum treatment
of the model will be given later.

The two-dimensional dilaton gravity model is described by the action
\be
\label{act}
S = {1\over 2\pi} \int \!
d^2 x \sqrt{-g} \Bigl[ e^{-2\phi} (R +4 (\nabla \phi)^2 + 4\lambda^2 )-
\half \sum_{i=1}^N (\nabla f_i)^2 \Bigl].
\ee
Here $\phi$ is the dilaton, which forms together with the metric
$g = ds^2$ the gravitational sector of the model, and the $f_i$ are
free massless matter fields. The classical and semi-classical properties
of this model have been the subject of many recent papers \cite{cghs},
\cite{bietal}-\cite{hs}, so we will only give a brief description here.

The classical equations of motion of (\ref{act}) can be integrated exactly
for arbitrary in- and out-flux of energy. The general solution describing
the gravitational collapse of an incoming amount of massless matter forming
a black hole is given by
\ba
\label{rhofi}
ds^2 \is e^{2\rho} dx^+ dx^-, \qquad \qquad \nonu
e^{-2\rho}\, = \, e^{-2\phi} \is -\lambda^2 x^+[x^- + {1\over \lambda^2}
P_+(x^+)] + M(x^+).
\ea
Here $P_+(x^+)$ and $M(x^+)$ are expressed in terms of the incoming
energy-momentum flux $T_{++}(x^+)$ as
\ba
P_+(x^+) \is \int^{x^+}_0 \! dy^+ T_{++}, \\[2.5mm]
\label{mass}
M(x^+) \is \int^{x^+}_0\! dy^+ y^+ T_{++}.
\ea
The form of this geometry is depicted in fig 1a for the case in which
the incoming matter pulse is localized in the form of an (approximate)
shockwave. The solution is obtained by gluing together, along the shockwave
trajectory, the linear dilaton vacuum solution
\be
\label{ldv}
e^{-2\rho}\, = \, e^{-2\phi} \, = \, -\lambda^2 x^+ x^-
\ee
for small $x^+$, to a static black hole solution \cite{2dbh,withole}
for large $x^+$
\be
\label{bhv}
e^{-2\rho}\, = \, e^{-2\phi} \, = \, -\lambda^2
x^+(x^- + {1\over\lambda^2}\PP_+) + M
\ee
with $\PP_+ = \PP_+(\infty)$ the total incoming Kruskal momentum and
$M = M(\infty)$ the total energy of the incoming matter wave.
The matching condition  that determines the form
of this solution is that the dilaton and metric are continuous along
the shockwave. Classical trajectories of out-going massless particles
are therefore straight lines without any discontinuity at the shockwave.

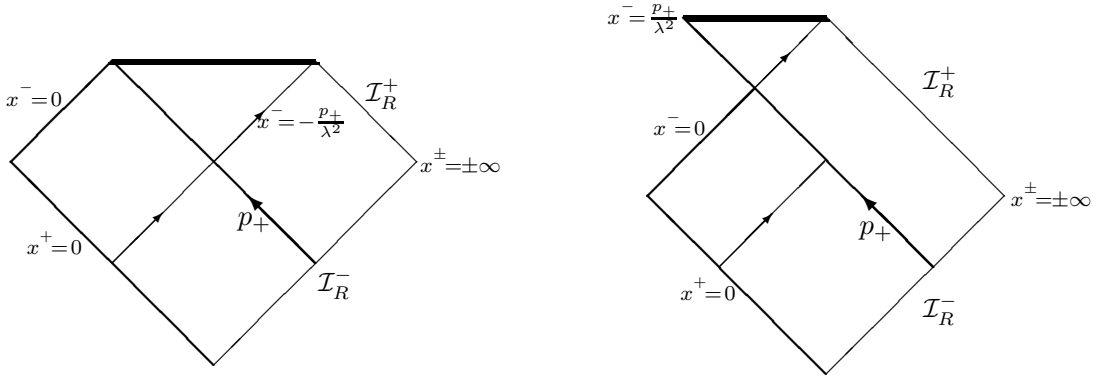
\figuur{1}{Figure 1a and 1b.
Two representations of the same classical geometry describing the
shockwave and subsequent black hole formation. In the left figure the
dilaton and metric are continuous across the shockwave, and rightmoving
lightrays are straight lines. On the right the lightray
trajectories undergo a discontinuous shift when crossing the shockwave.
The horizon is indicated in both figures as the last ray that reaches
${\cal I}_R^+$.}

Observers that do not decide to throw themselves into the black hole
have only access to that part of the space-time which is outside of
the horizon. Since we will in the following only be interested in
what these outside observers can see, it will be convenient to redefine
the $x^+,x^-$ coordinate system such that the observable part of space
time is always given by the region $x^+>0$ and $x^-<0$. In these
coordinates, the geometry describing the formation of a black hole
looks as in fig 1b. This geometry becomes physically equivalent to the
previous one, provided we take into account that the $x^-$ coordinate
of a right-moving particle undergoes a shift
\be
\label{shift}
\qquad x^- \rightarrow x^- +\delta x^- \qquad \qquad
\delta x^- = {1\over \lambda^2} \PP_+
\ee
when it crosses the shockwave trajectory. Thus particles for which initially
$x^->- \lambda^{-2} \PP_+$ will get shifted to positive $x^-$ values,
\ie \ inside the black hole region, and thus become invisible to outside
observers.

Now let us consider the effect of this shockwave on the quantum mechanical
wave function of the rightmoving $f$-particles. The coordinate shift
(\ref{shift}) is associated to the (formally unitary) quantum operator
\be
U = \exp( {i\over 2\pi} \delta x^- \PP_-)
\ee
where $\PP_- = \int_{-\infty}^0 dy^- T_{--}$ is the right-moving total Kruskal
momentum. After inserting that $\delta x^- = \lambda^{-2} P_+$, this
operator takes a form that is symmetric between the in-going
and out-going fields, as \cite{peetal}
\be
\label{ss}
U = \exp\Bigl({i\over 2\pi \lambda^2} \PP_+ \PP_-\Bigr).
\ee
So far, this quantum operator just represents the shockwave interaction
between the left- and right-moving $f$-fields.

A qualitative difference between two-dimensional dilaton
gravity and, for example, the spherical symmetric reduction of
Einstein gravity is that the linear dilaton vacuum does not provide
a natural (timelike) boundary analogous to the origin $r=0$. On the
other hand, the effective coupling constant $\kappa = e^\phi$ becomes
infinite near the left null boundaries of the linear dilaton vacuum,
and it is therefore not appropriate to treat these as asymptotic
regions. Instead it is more natural to try to impose suitable
boundary conditions in this strong coupling regime, which will
then prescribe the initial conditions for the right-moving fields.
An important consequence of this way of setting
up the model is that {\it initial data only need to be specified in the
right in-region.} The black hole information problem can be formulated
as the question whether (there exists a physically reasonable choice of
boundary conditions such that) all the initial
information will be contained in the final data that arrive in the right
out-region.

In the presence of a reflecting boundary condition, the operator
(\ref{ss}) describing the interaction between
the left- and right-moving modes acquires a new meaning. To
illustrate this, let us for the moment imagine that the matter
is described by only one single quantum mechanical particle.
The operator $\PP_+$ then measures the incoming Kruskal momentum of this
particle, and is conjugate to its coordinate $x^+$.
Similarly, $\PP_-$ measures the out-going momentum conjugate to $x^-$.
Now, because of the reflection condition, we are asked to identify
the in- and out Hilbert spaces, and this must be done in a way that
takes into account the shockwave interaction between the particles. This is
achieved if, following \cite{thooft},
we promote the operator (\ref{ss}) to the $S$-matrix relating
the in-state and the out-state
\be
\label{sss}
\langle \PP_+, out | \PP_-, in \rangle =
\exp\Bigl({i\over 2\pi \lambda^2} \PP_+ \PP_-\Bigr) \ .
\ee
This one-particle $S$-matrix is the dimensional reduction
of the black hole scattering matrix proposed by 't Hooft in
3+1-dimensions \cite{thooft}.

The above formula implies that, as a quantum operator,
the out-going coordinate $x^-$ is identified with the incoming momentum
operator via
\be
\label{xP}
\lambda^2 x^- =  \PP_+ \ .
\ee
This relation may look somewhat mysterious, but note that
classically, as seen in fig. 1b, this is precisely where the in-coming
particle ends up!
It leads, however, to an important difficulty with the interpretation
of (\ref{sss}) as an $S$-matrix. Namely, while (\ref{ss})
indeed looks like a unitary
operator in the single particle Hilbert space, this is true only if the
Hilbert space contains all eigenstates of $\PP_\pm$.
If, however, we are interested in what an outside observer sees, we should
restrict ourselves to in- and out-states that have support only
in the outside region $x^+ > 0 $ and $x^-<0$, and after this restriction
the $S$-matrix (\ref{sss}) is no longer unitary.
For example, a basis of states that has support only in the outside region
is provided by the eigenstates of the asymptotic energy operator $M$,
given by
\ba
\langle x^+|\omega, in\rangle \is
(\lambda x^+)^{-i\omega-\frac{1}{2}} \theta(x^+) \ ,\nonu
\langle x^-|\omega, out\rangle \is
(-\lambda x^-)^{i\omega-\frac{1}{2}} \theta(-x^-) \ .
\ea
The $S$-matrix is diagonal on this basis, with matrix elements (up to a phase)
\be
\langle \omega', out | \omega, in \rangle =
{1\over \sqrt{2\pi}} e^{-\frac{\pi}{2}\omega}\,
{\Gamma(\half-i\omega)}
\delta_{\omega,\omega'}\ .
\ee
The fact that these matrix elements are not phase factors,
but instead satisfy $|S|^2<1$, shows that
part of the wavefunction has disappeared behind the horizon.
This is of course no
surprise, since classically an in-going particle will never reach
${\cal I}^+_R$. The surprise is in fact that in quantum mechanics
a part of the final wavefunction does end up in the right out-region.
The form of this asymptotic out-state follows in an essentially unique
way from the shockwave interaction described above.

It is evident that (\ref{sss}) cannot be used as an $S$-matrix
in a second quantized model, as it only knows about
the total Kruskal momentum $\PP_\pm$.\footnote{It should be mentioned
that the above dimensional reduction is of course a slightly
misleading caricature of 't Hooft's $S$-matrix, as in 3+1-dimensions
the Kruskal momenta depend on the angular coordinates.} Instead, if
one would naively generalize the above semi-classical reasoning to the
second quantized theory, one will almost inevitably end up with the standard
conclusion that the out-going state becomes a mixed state that only depends
on the total incoming mass $M$ and momentum $\PP_+$  (see section 4.2).
The fact that for first quantized matter a part of the wave-function
disappears behind the horizon, would translate in second quantized language
to loss of coherence. However, in this reasoning one assumes that the only
interaction between the in- and out-fields is described by (\ref{ss}).
This is in fact not true: the fields also communicate with each
other via the reflecting boundary condition, and since this interaction
takes place in the strong coupling region, it cannot be ignored or treated
semi-classically.

In the following we will define a simple model for this boundary
interaction, and taking its quantum effects into account, we will
construct a generalization of the above one-particle $S$-matrix that
does keep track of the full structure of the quantum states.
This $S$-matrix will contain (\ref{sss}) as an overall factor acting in the
zero-mode space of the dilaton gravity fields. As a consequence
it will have the same property that it is unitary only in an enlarged
Hilbert space that includes a sector that is unobservable from outside
the black hole. However, since this concerns only the zero-mode
sector, the asymptotic state in the right out-region will be
pure and will contain all initial information.

\newsection{Quantization of $N=24$ Dilaton Gravity}

We will now describe the free field formulation of
dilaton gravity coupled to $N=24$ massless scalar fields
given in \cite{us2}, see also \cite{kazama}.
This formulation will reveal a close correspondence with critical
open string theory, which we will further exploit in section 5.

\newsubsection{Free Field Formulation}

In the conformal gauge $ds^2 = e^{2\rho} du\,dv$, two-dimensional
dilaton gravity is described by the action
\be
\label{Sconf}
S={2\over \pi}\int \!dudv\, \Bigl[e^{-2\phi}
(2\partial_u\partial_v\rho-4\partial_u\phi\partial_v\phi + \lambda^2
e^{2\rho}) + \hf \sum_{i=1}^N \partial_u f_i \partial_vf_i\Bigr] .
\ee
In the following we
will choose the number of matter fields to be $N=24$, in which case
the quantization of the theory is most straightforward. In particular,
in this case the classical dilaton gravity theory of the $\rho$ and $\phi$
fields does not receive any one-loop corrections due to the conformal
anomaly. The gravitational and matter sector each separately define a
conformal field theory, with respective energy-momentum tensors
\ba
T^g_{uu} \is
(4\partial_u \rho\partial_u \phi-2\partial_u^2\phi)e^{-2\phi}\ , \\[2mm]
T^m_{uu} \is  \sum_{i=1}^N \half (\partial_u f_i)^2 \ .
\ea
These
energy momentum tensors each generate a Virasoro algebra of central charge
$c\! =\! 2$ and $c\! =\! N\! =\! 24$, respectively.
The coupling between the two sectors is described via the Virasoro constraint
\be
\label{vicon}
T^g_{uu} + T^m_{uu} = 0 
\ee
which supplements the equations of motion of (\ref{Sconf})
and ensures the general covariance of the combined theory.

The general solution to the classical equation of motion
of (\ref{Sconf}) can be parametrized in terms of pure left or right-moving
fields as follows
\ba
e^{2\rho-2\phi}\is\partial_u\X^+(u)\partial_v\X^-(v) \label{rhoXX}
\\[2.5mm]
\label{phiXX}
e^{-2\phi} \is -\lambda^2 \X^+(u) \X^-(v) + \Omega^+(u) + \Omega^-(v) \\[2.5mm]
f_i \is f^+_i(u) + f^-_i(v) \ .
\ea
The first equation shows that in the classical theory one can
choose special light-cone coordinates in which $\rho = \phi$. The
required conformal transformation  $(u,v) \ra (\X^+,\X^-)$
are dynamical variables in the quantum theory.
The only mechanism by which the left and right-moving
fields interact is via the boundary conditions in the strong coupling regime.
These boundary conditions are necessary to
prescribe the initial conditions for the right-moving fields in terms
of the incoming fields, but also to effectively implement the restriction
on the chiral fields that follows from the condition that $e^{-2\phi} >0$.

To set up the quantum theory, we can therefore first concentrate on the
two chiral sectors separately, and impose the boundary conditions afterwards.
In the following we will only write the formulas for the left-moving fields.
It is convenient to introduce the field variable $\PP_+(u)$ via
\be
\label{opx}
\partial_u \Omega^+ = \PP_+ \partial_u\X^+ \ .
\ee
In this parametrization the gravitational energy-momentum
tensor takes the simple form
\be
T^g_{uu} = \partial_u \PP_+ \partial_u \X^+ \ .
\ee
In the quantum theory, this operator should generate conformal
transformations $u\ra \tilde{u}(u)$ on the fields $\X^+$ and $\PP_+$.
Hence it is a reasonable procedure to identify $\PP_+$ with the canonical
conjugate to the chiral coordinate $\X^+$, and define the quantum theory by
postulating the commutation relation
\be
[\partial_u\PP_+(u_1), \X^+(u_2)] = - 2\pi i \delta(u_{12}),
\ee
where $u_{12} = u_1 -u_2$.
In this way we arrive at a free field formulation of the pure dilaton gravity
theory away from the boundary.

Because of the somewhat unconventional asymptotic conditions on the $\X$
fields, we can not simply use the standard mode-expansion. We find that
the only mode-expansion that is consistent with the required asymptotic
behaviour is of the form
\newcommand{\Xmode}{\mbox{\sl x}}
\newcommand{\Pmode}{\mbox{\sl p}}
\ba
\label{xpans}
\partial_u \X^+(u)\is {x^+e^{\lambda u}} \ +
{e^{\lambda u} }\int d\omega \, \Xmode^+(\omega)\, e^{-i\lambda \omega u}\ ,\\
\partial_u \PP_+(u)\is {p_+e^{-\lambda u}} \ +
{e^{-\lambda u} }\int d\omega \, \Pmode_+(\omega)\, e^{-i\lambda \omega u}\ ,
\ea
where $x^+$ and $p_+$ are $c$-numbers and the other
modes satisfy the algebra
\be
[\Xmode^+(\omega_1),\Pmode_+(\omega_2)] =   (\omega_1 + i)
\delta(\omega_1 \! + \! \omega_2)\ .
\ee
The $c$-number coefficients in front of the
first terms in (\ref{xpans}) can be fixed by requiring that
the conformally normal ordered energy-momentum tensor $T^g_{uu}$ has no
vacuum expectation value. One finds that this implies the condition
\be
x^+p_+ = -{\lambda^2 \over 2} \ .
\ee

A possible criticism of the above quantization procedure is that,
by treating the dynamical coordinates $\X^\pm$ as
free fields, we have given up the restriction that they should
be invertible functions of $u$ and $v$, respectively. This could create some
potential problems with the correspondence principle. However, we are helped
here by the fact that the quantum dynamics of the whole model is
constrained by the Virasoro conditions (\ref{vicon}) and thus invariant
under conformal transformations. This invariance can be used, if one wants,
to choose a light-cone gauge and fix $\X^+$ or $\X^-$ to be specific regular
functions of the coordinates.
One thereby ensures, at least asymptotically, that no
degenerations occur. In the strong coupling region, however,
large quantum fluctuations can still lead to possible acausal behaviour,
but it seems a reasonable assumption that this does not lead to
unacceptable physical consequences in the classical region.

\newsubsection{Physical Operators}

As in any theory of quantum gravity, only operators that have a
coordinate invariant definition correspond to physical observable
quantities. In the quantum theory this corresponds to the requirement
that the operators must commute with the Virasoro conditions.
The most convenient way to obtain these physical
operators is by using the dynamical fields $\X^\pm$ as a reference
coordinate system. This procedure corresponds to choosing the gauge
$\rho=\phi$. It is further important to note that the asymptotic
light-cone coordinates defined by the physical metric
are given by $\tau^\pm = \pm \log(\lambda\X^\pm)$. These coordinates are
what asymptotic observers identify as their proper time, and we must
therefore use them to define physical quantities such as energy, etc.

In the left-moving sector, we can thus associate a local physical operator
to each $f_i$-field as follows
\be
\label{fin}
f^{(in)}_i(x^+)
= i \int \frac{d\omega}{ \omega} (\lambda x^+)^{-i\omega} \alpha_i(\omega)
\ee
where $x^+$ is a $c$-number and  $\alpha_i(\omega)$ the operator
\be
\label{Ao}
\alpha_i(\omega) = {1\over 2\pi} \int\! du\,
        (\lambda \X^+(u))^{i \omega} \partial_uf_i(u) .
\ee
These operators $\alpha_i(\omega)$ manifestly
commute with the Virasoro operators,
and thus create/annihilate physical modes of the $f$-field of energy
$\omega$ (measured in units of $\lambda$). They satisfy the algebra
$[\alpha_i(\omega_1),\alpha_j(\omega_2)] = \omega_1 \delta_{ij}
\delta(\omega_1+\omega_2)$, and
are the direct analogues of the DDF operators that span the
physical Hilbert space in critical string theory \cite{ddf}.
We can similarly define a complete basis for the right-moving sector via
\be
\label{fout}
f^{(out)}_i(x^-) = -i \int {d\omega\over \omega}
(\lambda x^-)^{i\omega} \beta_i(\omega)
\ee
where $x^-$ is a $c$-number and  $\beta_i(\omega)$ the operator
\be
\label{Bo}
\beta_i(\omega) = {1\over 2\pi}
\int \! dv\, (-\lambda \X^-(v))^{-i \omega} \partial_vf_i(v) .
\ee
The fact that these physical in- and out-modes
have such a simple expression in terms of the
fundamental field variables is a major advantage of the present
formulation of dilaton gravity.

In the following it will be crucial to know for sure
that the modes $\alpha_i$ and $\beta_i$ indeed define a complete basis
of the left- and right-moving sectors of the physical Hilbert space.
In particular, this would mean that any other physical quantity in our
model can be expressed in terms of these modes.
In section 5 we will show that this is indeed the case.
As an example, let us discuss here the
physical operator associated with the dilaton field $\phi$ at
a given point $(x^+,x^-)$ in the $\rho = \phi$ coordinate system.
Using the field redefinitions (\ref{phiXX}) and (\ref{opx}), we
find that this operator is expressed in terms of the $\X$ and $\PP$
variables as follows
\be
\label{sol}
e^{-2\phi}(x^+,x^-) = M -
\lambda^2 x^+ x^- - \int_{{x^+}}^{{}^{\!\infty }}\!\!\!
dy^+ \PP_+(y^+) - \int^{x^-}_{{}_{ \! - \infty }}\!\!\!\!
dy^- \PP_-(y^-) \ ,
\ee
where the constant $M$ is identified with the black hole mass
and e.g. $\PP_+(x^+)$ is defined via
\ba
\PP_+(x^+) \is - \int {d\omega\over 1+i\omega} (\lambda x^+)^{-1-i\omega}
\hat\PP_+(\omega) \ ,
\nonu
\label{hatpo}
\hat{\PP}_+(\omega) \is  {1\over 2\pi} \int \! du \,
(\lambda \X^+)^{1+i\omega} \del_u \PP_+(u) \ .
\ea
In the quantum theory,
the composite operator in the integral in (\ref{hatpo})
has to be normal ordered in a conformally invariant fashion.
The precise procedure will be described in section 5. The resulting
operators $\hat{\PP}_+(\omega)$ are then physical operators, analogous
to the operators introduced by Brower in his proof of the no-ghost
theorem in string theory \cite{brower}.
{}From their form (\ref{hatpo}) it can be seen that they generate
diffeomorphisms of the physical coordinate $\X^+$, and it can indeed be
verified that they satisfy a Virasoro algebra with central charge $c=24$
\cite{brower}. These two facts are sufficient to show that within the
BRST-cohomology ({\it i.e.}\ modulo spurious physical fields) the field
$\PP_+(x^+)$ is identified with the integral
\be
\label{ppt}
\PP_+(x^+) = \int_{x^+}^{{}^{\!\infty}} \!\!\!
dy^+ T_{++}(y^+)
\ee
of the left-moving physical energy-momentum tensor
\be
T_{++}(x^+) = \hf \sum_{i}(\partial_+ f_i(x^+))^2 \ ,
\ee
which can be expressed directly in terms of the $\alpha_i(\omega)$.
A similar formula holds for $\PP_-(x^-)$.

We have thus found that the classical statement that the matter energy
momentum flux uniquely determines the form of the dilaton and the metric,
as expressed by the eqns.\ (\ref{rhofi}) - (\ref{mass}),\footnote{Note that
the coordinate $x^-$ has been shifted with an amount $\PP_+/\lambda^2$.}
has led to
the quantum identification (\ref{ppt}) of operators within the physical
Hilbert space. Due to this, a state in the physical Hilbert space is
uniquely characterized once we specify its in-going and out-going matter
content, so that the $\alpha_i$ and $\beta_j$-modes each form a complete
basis for the physical Hilbert space.
The $\alpha_i$ and $\beta_j$-operators
are independent as long as we do not impose any reflection condition
at strong coupling. However, we will now formulate the boundary condition
that will lead to an identification
between the left- and right-moving Hilbert space sectors and thus to a
relation between both types of physical modes.

\newsubsection{The boundary condition}

Since we would like to set up the
model in such a way that initial data need to be specified only in the right
in-region, we must introduce some reflection condition in the strong
coupling regime to prescribe the initial conditions for the right-moving
fields. However, since the classical model is unstable against
black hole formation, it is a non-trivial problem to find
a boundary condition that satisfies all reasonable physical requirements.
Namely, on the one hand it should lead to a consistent quantum model, while on
the other hand is should not lead to major modifications of the classical
physics. In combination, these two requirements are very restrictive.

Perhaps the most natural procedure is to pick a line
on which the dilaton $\phi$ takes a large but finite value, and define this
line to be a reflecting boundary. In the vacuum this line of constant $\phi$
is of the form $\lambda^2 x^+x^- = -\epsilon$, with $\epsilon$ some
infinitesimal positive constant.
In general, however, the boundary trajectory is more complicated, since it
depends in a dynamical
way on the incoming energy flux. We will now define a simple model
for the boundary dynamics, which will enable us to take this
backreaction into account quantum mechanically. The main justification
for the specific choice we will make is that it is the simplest one
that appears to lead to the correct physics to an outside observer.
In particular, as we will see, it preserves the property that a black hole
can be formed.

Now let us describe the boundary condition. (The following boundary
condition was first proposed and studied for large $N$ dilaton gravity
by Russo, Susskind and Thorlacius \cite{rst}.)
We choose the $(u,v)$-coordinate system in such a way that
the boundary becomes identified with the line $u=v$, and denote
the parameter along this boundary by $s$. As suggested above, we
first require that the dilaton field is
constant along the boundary. In terms of the $\X$ and $\PP$-fields this
condition reads
\be
\partial_s e^{-2\phi}  =
-(\lambda^2 \X^- - \PP_+)\partial_s \X^+  -
(\lambda^2 \X^+ + \PP_-)\partial_s \X^- = 0 \ .
\ee
In addition we will require that the incoming energy momentum
flux gets directly reflected off the boundary. We demand this for the
gravitational and matter components separately
\ba
\partial_s \X^+ \partial_s\PP_+ \is -\partial_s \X^- \partial_s \PP_-\nonu
\sum_i \hf (\partial_s f_i^+)^2 \is \sum_i \hf (\partial_s f_i^-)^2 .
\ea
The simplest  (but certainly not unique)
way to ensure that all these boundary conditions are satisfied
is to make the following identifications
\ba
\label{reflect}
\qquad \qquad {\lambda^2}\X^- \is \PP_+  \nonumber\\
\qquad \qquad {\lambda^2}\X^+  \is   -\PP_-  \qquad
\qquad {\mbox\rm at} \ u=v \, . \\
\qquad \qquad f^+ \is f^- \nonumber
\ea
These equations should be read as quantum identifications between
operators in the left- and right-moving sector of the theory.
In the remainder of this paper we will adopt
these reflection conditions and study their physical consequences.

An unusual property of the above reflection equations (\ref{reflect})
is that they do not have an obvious classical interpretation.
In the first place, since
the boundary lies in the strong coupling regime, strong quantum fluctuations
will produce a large uncertainty in its precise location.
Still, one can try to describe the reflection process semi-classically by
first solving the boundary trajectory in terms of the incoming energy
flux and then determining the out-going fields via direct reflection
off it. This boundary trajectory can be specified as a
relation between the coordinates $\X^\pm$, which from (\ref{reflect})
and (\ref{ppt}) is seen to take the form
\be
\label{xxt}
\lambda^2\X^-(x^+) = \int_{x^+}^{{}^{ \infty}} \!\!\!
dy^+ T_{++}(y^+) \ .
\ee
Combined with the reflection equation
\be
f^{(in)}(x^+) = f^{(out)} (\X^-(x^+))
\ee
this procedure leads to an explicit non-linear scattering
equation between the in- and out-going fields \cite{us2}.
This scattering equation, however, will be of a rather unusual type.
The incoming energy-flux $T_{++}$ is classically always greater or equal
to zero, which means that the boundary trajectory (\ref{xxt}) is always
space-like or light-like. Moreover, the right-hand side of (\ref{xxt})
is always positive, whereas the out-going fields are supposed to
come out at negative $x^-$. The identifications (\ref{reflect})
can therefore not simply be interpreted as a direct classical reflection
of some time-like boundary. Instead, they show that the classical
boundary trajectory always stays behind the global event horizon, and thus
its presence does not modify the properties of the classical theory
as seen by an outside observer. In a way, this is what we want, because
it shows that we are still dealing with a theory in which infalling matter
leads to black hole formation.

To get a better idea of the physical role of the boundary
condition, let us for the moment
imagine that the energy momentum flux $T_{++}$ in (\ref{xxt})
contains an infinitesimal negative `vacuum' contribution
$\langle T_{++}\rangle = -\epsilon/(x^+)^2$.\footnote{In fact, semi-classical
arguments \cite{rst,us2} suggest that the quantum energy momentum
tensor $T_{++}$ indeed receives a constant negative contribution
from vacuum fluctuations. We will
comment further on the role of this vacuum energy in section 4.3.}
The line (\ref{xxt}) will then be time-like in the vacuum, and thus
the reflection will in particular
identify the left- and right-moving vacuum states.
During collisions with the incoming particles, however, the boundary
trajectory will still become
space-like, and from this point on the classical description of
the theory breaks down. Indeed, in the classical dilaton gravity
model, the boundary then goes over in a black hole singularity.
In our parametrization of the dilaton gravity fields,
what happens instead is that the mapping from the $(u,v)$ parameter
space to the physical coordinate space $(\X^+,\X^-)$ becomes
non-invertible. If we assume that the $\X^+$ is a monotonic function of $u$,
then it is easy to show that the $\X^-$ coordinate will turn around
at the point where normally the singularity is formed and go backwards
in time for a while. This implies that, in our model, the propagation
in the physical coordinate space can become acausal near the singularity.

\begin{center}
\leavevmode
\epsfysize=6cm \epsfbox{fig2.ps}
\end{center}

\begin{center}
{\parbox{15cm}{\small  Fig 2.
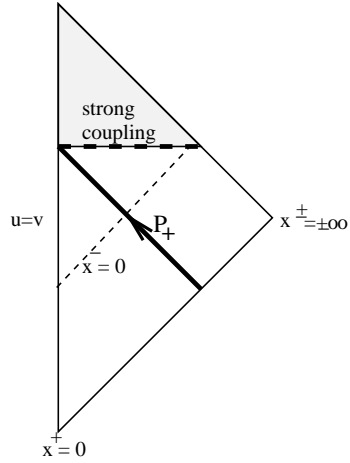
The complete black hole formation process takes place within the
region of the $(u,v)$ plane where the fields $\X^\pm$ provide an
invertible parametrization. The acausal strong
coupling effects take place in the shaded region, \ie\ near the
black hole singularity.}}
\end{center}

It is important to note, however, that up to this point
our model is classically completely equivalent to the standard dilaton
gravity theory. In particular, as indicated in fig. 2, it can be seen
that the complete black hole formation process takes place within the
region of the $(u,v)$ plane where the fields $\X^\pm$ are
non-degenerate, invertible functions.
The acausal strong coupling effects take place in the shaded region,
which is after the black hole is formed, and do not modify the  classical
physics as seen by an out-side observer.
The present model should therefore not be viewed as a {\it modification}
of the original dilaton gravity theory, but rather as a {\it completion}
of it, in which a certain {\it Ansatz} is made about what happens near the
singularity. This {\it Ansatz} is motivated and strongly constrained by
the observable properties of the model.

Finally, it is instructive to make a comparison between the above model
and the two-dimensional reduction of 't Hooft's black hole $S$-matrix
discussed in section 2. Indeed, the operator identifications (\ref{reflect})
are the simplest generally covariant generalization of the relation (\ref{xP})
between the out-going coordinate $x^-$ and in-going momentum $\PP_+$
of the matter particle considered there. In that case we saw that,
although (\ref{xP}) classically states that the particle ends up behind
the event horizon, quantum mechanically it uniquely predicts the form of the
out-going wavefunction on ${\cal I}_R^+$. In the following sections
we will show that the identification (\ref{reflect}) leads in the
same way to a unique out-going wavefunction in the full dilaton gravity
theory.

\newsection{Physical Properties of the $S$-matrix}

The reflection equations (\ref{reflect}) prescribe the initial conditions
for the right-moving modes by identifying the right- and left-moving sectors
of the dilaton gravity Hilbert space. Thus we can now describe the full
Hilbert space in terms of, say, only the left-moving matter fields
$f_i(u)$ and the coordinate fields $\X^\pm(u)$, which
become each others canonical conjugate
\be
[\partial_u \X^\pm(u_1), \X^\mp(u_2)] = -{2\pi i\over \lambda^2}
\delta(u_{12}).
\ee
This allows us to define the $S$-matrix elements
between physical in- and out-states as an expectation value
of the corresponding product of in- and out-creation- and
annihilation operators
\be
\label{ab}
\langle out | in \rangle =
\langle 0 | \prod_m \beta_{j_m}(\omega_m)
      \prod_n \alpha_{i_n}(\omega_n) |0\rangle
\ee
where $\alpha_{i}$ and $\beta_j$ are defined in (\ref{Ao})-(\ref{Bo}).
Because the $\alpha_i$ and $\beta_j$ modes each form a complete basis
of the physical Hilbert space, the collection of $S$-matrix elements
(\ref{ab}) define a unique and invertible mapping from a given in-state to
an out-state. A more precise algebraic description of this mapping will be
given in section 5. The purpose of this section is to investigate the
physical properties of the above $S$-matrix. We will show that it indeed
largely behaves as one would expect from semi-classical considerations.
This correspondence provides an important justification for our simple
choice of boundary condition.

\newsubsection{The Exchange Algebra.}

The observation that will enable us to make contact with
semi-classical results obtained in previous studies of dilaton gravity
\cite{cghs}\cite{rst}
is that the interaction between the in- and out-modes can be naturally
separated into two parts, namely the shockwave interaction
discussed in section 2 and the reflection off the
dynamical boundary. The key property that distinguishes
the shockwave interaction from the other interactions is that
it already affects
the form of the out-state before the incoming particles have reflected.
Furthermore, as we will see,
it is directly responsible for the presence of thermal radiation
in the asymptotic out-state.
The other interactions, describing the
reflection, will of course lead to important corrections to the Hawking
spectrum and are responsible for restoring the coherence of the out-state.

A natural but naive attempt to compute the expectation value (\ref{ab})
is to try to commute the operators $\beta_j$ through the operators
$\alpha_i$, until they annihilate the opposite vacuum. This approach
suggests that it may be useful to summarize the interaction between
the in- and out-modes $\alpha_i(\omega)$ and $\beta_j(\omega)$
in terms of an exchange algebra. Now it is well-known from conformal
field theory that
the two vertex operators $e^{-{i\over 2\pi}p_+X^+}$ and
$e^{-{i\over 2\pi}p_-X^-}$ indeed
have simple exchange-properties: if we interchange their order one simply
picks up a phase that only depends on the sign of their relative
position\footnote{Here we used
that $[X^-(u_1),X^+(u_2)] = 2\pi i\lambda^{-2} \theta(-u_{12})$.}
\be
\label{sexch}
e^{-{i\over 2\pi}p_- X^-(u_1)}
e^{-{i\over 2\pi}p_+ X^+(u_2)}
= e^{-{i \over 2\pi \lambda^2}p_+p_-\theta(-u_{12})}\,
e^{-{i\over 2\pi}p_+ X^+(u_2)}
e^{-{i\over 2\pi}p_- X^-(u_1)}.
\ee
We would now like to use the above formula (\ref{sexch})
to determine an exchange algebra of the physical $in$ and $out$ $f$-fields.

To this end, let us first introduce the following physical operators
\ba
\label{ABhat}
\widehat{A}_i(p_+)\is {1\over 2\pi}\int\! du \,
e^{-{i\over 2\pi}p_+X^+(u)}\partial_uf_i(u)\nonu
\widehat{B}_j(p_-)\is
{1\over 2\pi}\int\! du\, e^{-{i\over 2\pi}p_-X^-(u)} \partial_uf_j(u),
\ea
which create in-coming and out-going particles with a definite Kruskal-momentum
$p_+$ and $p_-$ respectively. These operators can be expressed
as linear combinations of the energy eigenmodes
$\alpha_i(\omega)$ resp. $\beta_j(\omega)$. For example, for $p_+>0$ we have
\be
\label{Alpha}
\widehat{A}_i(p_+) = \int\! d\omega\, e^{-{\pi\over 2}\omega}
 \Gamma(-i\omega)
 \Bigl({p_+\over2\pi \lambda}\Bigr)^{i\omega} \alpha_i(\omega).
\ee
This equation, which can also be read as the definition of
$\widehat{A}_i(p_+)$, shows that these Kruskal modes are in fact somewhat
singular operators, because
they contain $\alpha_i(\omega)$ modes of arbitrarily
high frequency. In the following we will mostly ignore this singularity,
as it will not affect the main conclusions.

The Kruskal modes (\ref{ABhat}) are very convenient for our purposes, since
from (\ref{sexch}) we find that they satisfy an exchange algebra of the
following form
\be
\label{pk}
\qquad \widehat{B}_j(p_-)\widehat{A}_i(p_+)
=e^{-{i \over 2\pi \lambda^2}p_+p_-} \widehat{A}_i(p_+)
\widehat{B}_j(p_-) + R_{ij}(p_+,p_-).
\ee
Here in the first term on the right-hand side
we recognize the two-particle $S$-matrix described in section 2,
representing the shockwave interaction between the in- and out-mode.
The remaining term takes the form
\ba
\label{rij}
R_{ij}(p_+,p_-) \is \int \! du_1 du_2  e^{-{i\over 2\pi}p_-X^-(u_1)}
e^{-{i\over 2\pi}p_+X^+(u_2)} \,
\Bigl[
2\pi i \delta_{ij}\delta^\prime(u_{12}) \nonu \qquad \qquad  \qquad \qquad
& &  \qquad \qquad \qquad  \qquad
\; + (1-e^{-{i \over 2\pi \lambda^2}p_+p_-})f'_i(u_2) f'_j(u_1)
\theta(u_{12})\Bigr],
\ea
and can be seen to describe the effects due to the presence of the dynamical
boundary. Namely, the integrand
in (\ref{rij}) has support only for
$u_1\geq u_2$ and this operator $R_{ij}(p_+,p_-)$ therefore only contains
interactions that take place at the moment of reflection or afterwards.
The first term in the integrand represents the direct reflection off
the boundary, while the second term corrects for the fact that
the shockwave interaction between the in and out-modes takes place
only when the in- and out-mode cross each other before the reflection
has taken place.

The physical interpretation of (\ref{pk}) becomes a little
more clear when we go to
a coordinate representation for the outgoing modes.
In terms of the coordinate field $f^{(out)}(x^-)$, defined in (\ref{fout}),
the exchange algebra takes the form
\be
\label{shft}
\qquad f_j^{{(out)}}(x^-) \, \widehat{A}_i(p_+) =
 \widehat{A}_i(p_+) \, f_j^{{(out)}}(x^-\!\! -\!{p_+\over \lambda^2}) +
\tilde{R}_{ij}(p_+,x^-),
\ee
and we explicitly see that the in-mode shifts the out-fields by an
amount proportional to the incoming Kruskal-momentum.
The reflection term
$\tilde{R}_{ij}(p_+,x^-)$ is the $x^-$ Fourier transform of (\ref{rij}).
By construction, this term only affects the form of the out-going wave
function after the incoming wave has reflected.

\newsubsection{Hawking radiation.}

We would like to use the above results to obtain some physical insight
into what the out-state will look like for a given instate.
Specifically, we wish to consider an incoming state in which the
matter is localized in a finite time interval
$x^+_0<x^+<x^+_0+\Delta x^+$ and carries a large total
energy $E \pm \Delta E$. This in-state may be represented
as a sum of eigenstates of the total Kruskal-momentum with
eigenvalues are concentrated around $\PP_+=E/x^+_0$.
Before these incoming high energy modes will reflect off the boundary,
they will first interact with the outgoing fields via the gravitational
shockwave. In this subsection we will study the physical effect of this
shockwave.

Since the incoming wave is localized, it is reasonable to assume
that the quantum reflection will take place within some finite outgoing
time interval concentrated around some reflection time $x^-_0$.
In the following we will be interested in the structure of the out-state
before this reflection time, which means that we are allowed to
ignore the presence of
the reflection term $R_{ij}(p_+,x^-)$. (A more detailed
justification for this procedure will be given in subsection 4.3.)
Thus, for the purpose of the following discussion, we are left with the
pure exchange algebra
\be
\label{shft2}
\qquad \qquad \qquad \qquad f_j^{{(out)}}(x^-) \, \widehat{A}_i(p_+) =
 \widehat{A}_i(p_+)\, f_j^{{(out)}}(x^-\!\! -\! {p_+\over \lambda^2}),
\qquad \qquad x^- < x^-_0 .
\ee
When the incoming energy $p_+$ is large enough, this
shift-interaction can produce large physical effects in the out-region.
Indeed, the shockwave due to an incoming particle can classically lead to
the formation of an event horizon, and, as is well known, the
resulting distortion of the out-modes produces Hawking radiation.
In the following we will rederive this result in our model, using
equation (\ref{shft2}) as a starting point. The derivation requires
only a small adaptation of the standard reasoning. Some
calculations done in \cite{gidnel} are also useful.

If the coordinate $x^-$ were a normal Minkowski coordinate, ranging
from $-\infty$ to $+\infty$, a constant shift in $x^-$ would
have had no physical effect whatsoever: it could simply be absorbed
by shifting the Minkowski vacuum. In our case, however, it is
crucial that $x^\pm$ parametrizes only a Rindler wedge $\pm x^\pm>0$
and that the vacuum of the $f$-fields is defined accordingly in
terms of the Rindler type modes $\alpha(\omega)$ and $\beta(\omega)$.
To see how the outgoing vacuum state is affected by the coordinate
shift (\ref{shft}), let us rewrite the exchange algebra in terms of the
$\beta$-modes.  One finds
\be
\label{exch}
\beta_j(\omega) \widehat{A}_i(p_+) =  \widehat{A}_i(p_+)\int\!d\xi
\,{\rm B}_{\omega\xi}(p_+)\beta_j(-\xi)
\ee\
with
\be
\label{bcoef}
B_{\omega\xi}(p_+) =
{1\over 2\pi}\Bigl({p_+\over\lambda}\Bigr)^{-i(\xi+\omega)}
{\Gamma(1-i\omega)\Gamma(i(\xi+\omega))\over\Gamma(1+i\xi)}.
\ee
The linear combination of $\beta$-modes
on the right-hand-side contains both creation- and
annihilation- operators. Consistency of the algebra (\ref{exch}) further
requires that these combinations again satisfy canonical commutation
relations, so we see that exchanging a $\beta_j(\omega)$-oscillator with
$\widehat{A}_i(p_+)$ leads to a Bogoliubov-transformation. Note that
the transformed modes occurring on the
right-hand side of (\ref{exch}) do not form a complete basis of all
$\beta$-modes, since they cover only the interval
$x^-<-\lambda^{-2} p_+$.  The exchange property (\ref{exch}) can
therefore in general only be used in one direction.

Now let us consider an in-state with definite total Kruskal momentum $\PP_+$,
and let us further assume it can be written in the form
\be
\label{psi}
|\psi\rangle =\prod_a \widehat{A}_{i_a}(p_a)|0\rangle,
\ee
with $\sum_a p_a=\PP_+$.
To determine the properties of the corresponding out-state
we can act on $|\psi\rangle$ with the $\beta_j$-modes and
repeatedly use (\ref{exch}) until we can act on the vacuum.
These manipulations are of course the direct quantum counterpart of
Hawkings original semi-classical calculation.
The repeated use of the exchange-algebra describes the propagation of the
outgoing particles through the infalling matter, while taking into account
the gravitational interaction between the two. Now,  it is clear  from
(\ref{shft2}) that in this procedure only the total
momentum $\PP_+$ plays a role, so the exchange relation between
the $\beta$-modes and the product $\prod_a\widehat{A}_{i_a}(p_a)$
is again of the same form (\ref{exch}). In this way we find that, just as in
the semi-classical calculation, the asymptotic out-state is no longer
equal to the vacuum-state, but given by the Bogoliubov transform
(\ref{exch})-(\ref{bcoef}) of the vacuum.

Although at this point we could simply refer to the standard
analysis \cite{hawking,gidnel},
let us continue to show that the spectrum of out-going radiation
indeed looks approximately  thermal. To this end, let us
compute the expectation value of the particle number
operator
\be
\langle N_j(\omega)\rangle ={1\over\omega}\langle \psi |
\beta^*_j(\omega)\beta _j(\omega)|\psi\rangle.
\ee
Inserting the definition of $|\psi\rangle$ we find that the right-hand-side
can be written as
\ba
\langle N_j(\omega)\rangle \is {1\over \omega}
\langle 0|\prod_b \widehat{A}^*(p_b)
\beta^*_j(\omega)\beta _j(\omega) \prod_a \widehat{A}(p_a)|0\rangle \nonu
\is \int_0^\infty\!d\xi \,{\xi \over \omega} |{\rm B}_{\omega\xi}(\PP_+)|^2,
\ea
where we made use of the exchange algebra (\ref{exch}),
and the fact that the vacuum-state is
annihilated by $\beta_j(\omega)$ for $\omega>0$. We further
assumed that the state $|\psi\rangle$ is normalized.
Applying some standard formulas about $\Gamma$-functions then gives
\be
\label{integral}
\langle N_j(\omega)\rangle= C\cdot\int_0^{\infty}\!\!
{d\xi\over \omega\!+\!\xi} \,
{\sinh \pi\xi\over \sinh \pi (\omega+\xi)
\sinh\pi \omega}.
\ee
It can be seen that
the dominant contribution in the integral comes from large $\xi$,
which allows us to approximate the integrand.
In this way we find that the expectation value of the particle number
operator is indeed given by a thermal distribution
\be
\label{therm}
\langle N_j(\omega)\rangle= C^\prime\cdot {e^{-2\pi\omega}\over
1-e^{-2\pi\omega}} \ .
\ee
The above reasoning can be extended in a straightforward way
to  analyze arbitrary expectation values of products of $\beta_j$-modes.

The integral in (\ref{integral}) as it stands would actually lead to
an infinite constant $C^\prime$ in (\ref{therm}), which would
imply that the out-state contains thermal radiation for arbitrary
late times. However, it is important to note that above we have of course
dealt with an idealized situation. In the first place we assumed that
$|\psi\rangle$ is an exact eigen state of $\PP_+$, and as
noted before, such states contain arbitrarily high energy modes and
infinite total energy. If the in-coming energy were bounded,
the resulting black hole would of course radiate for only a finite amount
of time. Moreover, after a certain time $x^-_0$ the reflection term $R_{ij}$
in the algebra (\ref{shft}) will become important and this
will produce important corrections to the Hawking spectrum. We will now discuss
the resulting modifications of the above semi-classical picture.

\newsubsection{Quantum Reflection.}

As discussed in section 3.3, the classical boundary trajectory always
stays behind
the global event horizon $x^- = 0$, and thus remains invisible to an outside
observer. In the quantum theory, however, fluctuations of the boundary will
be able to produce observable effects outside the horizon.
These effects are represented by reflection term $R_{ij}$
in the exchange algebra (\ref{shft}) between the in- and out-fields.
In the following it will be useful to write this algebra
in the coordinate representation
\be
 f_j^{(out)}(x^-) f_i^{(in)}(x^+) =
e^{{2\pi i\over \lambda^2} \partial_+ \partial_-}
f_i^{(in)}(x^+) f_j^{(out)} (x^-) + {\hat R}_{ij}(x^+,x^-)
\ee
where the reflection term ${\hat R}_{ij}(x^+,x^-)$ is the Fourier
transform of (\ref{rij}). We would now like to determine the regime
in which the presence of this term will become important.
The idea is to use the fact that the integrand in (\ref{rij})
vanishes for $u_2\geq u_1$ to show that the reflection term is negligible
to the right of some critical line in the physical $x^\pm$-plane.
For simplicity, we will restrict our discussion to only the first term
in (\ref{rij}). In coordinate space it reads
\be
\label{rij2}
\Delta_{ij}(x^+,x^-) = 2\pi i \delta_{ij} \int\! du \,
\delta(x^+\!\!-\!\X^+(u)) \partial_{u}\delta(x^-\!\!-\!\X^-(u))
\ee
where we performed one of the $u$-integrations compared to (\ref{rij}).
We will ignore the issue of normal ordering, as our discussion will be
semi-classical.

The above operator (\ref{rij2})
describes the direct reflection off the dynamical boundary,
and it will therefore be mainly responsible for the information transfer
from the in-going to the out-going modes. Using the first $\delta$-function
to perform the remaining integral over $u$, we can rewrite the right-hand
side of (\ref{rij2}) as
\be
\label{deltaij}
\Delta_{ij}(x^+,x^-) = 2\pi i \delta_{ij} \partial_+ \delta(x^- -
\X^-(x^+)) \ .
\ee
Here $\X^-(x^+)$ is the physical operator representing the boundary
trajectory in the $x^\pm$-plane, which by the equations of motion
is given in terms of the incoming energy flux $T_{++}$ via equation
(\ref{xxt}). From (\ref{deltaij}) we see explicitly that the operator
$\Delta_{ij}(x^+,x^+)$ indeed represents the direct reflection,
and furthermore, that it contributes only when its argument $(x^+,x^-)$
defines a point on the boundary line. In a similar way it can be shown
that the other term in $\hat{R}_{ij}(x^+,x^-)$ has its support behind
this same line.

So, from this we are led to conclude that physical effect of the
complete reflection term is negligible as long as we remain
to the right of the semi-classical boundary trajectory
\be
\label{estim}
\qquad \qquad \hat{R}_{ij}(x^+,x^-) \simeq 0
\qquad \qquad x^-  < \X^-(x^+),
\ee
with $\X^-(x^+)$ given in (\ref{xxt}). What we mean here by the
semi-classical boundary trajectory is the outermost line that the
boundary can reach when we include the effect of quantum fluctuations.
A semi-classical estimate of the magnitude of these quantum effects
can be given by considering the contribution of vacuum fluctuations to
the energy-momentum tensor $T_{++}$.
Namely, it is well-known that the definition of the quantum tensor
$T_{++}$ depends on the normal ordering prescription.
A natural choice is to normal order with respect to the local
Kruskal coordinates $x^\pm$, and in this case the physical
vacuum state will in fact contain a negative vacuum energy, given by
$\langle T_{++}\rangle = -{N\over 24 (x^+)^2}$.
Although we do not want to assign any real physical meaning to
this negative vacuum energy --- as it depends on the choice of normal ordering
--- it does give an indication of how large an effect
quantum fluctuations can have on the position of the boundary.

\begin{center}
\leavevmode
\epsfysize=6.5cm \epsfbox{fig3.ps}
\end{center}

\begin{center}
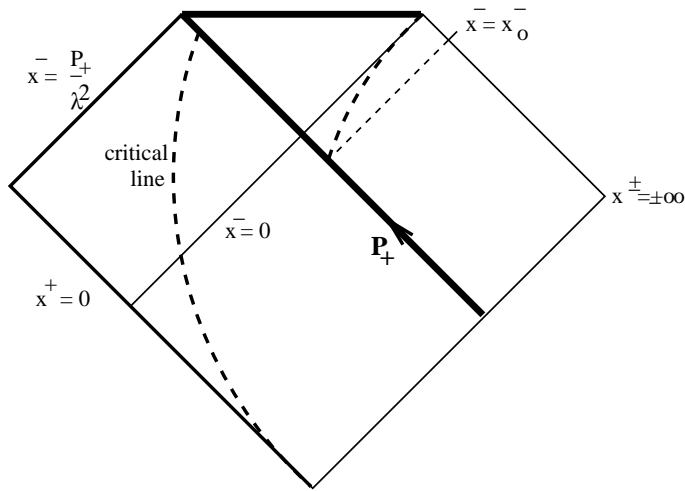
{\parbox{15cm}{\small  Fig 3. Again the collapsing black hole geometry,
where we have now indicated the critical line behind which the quantum
effects of the boundary can become visible. These quantum effects give
corrections to the Hawking spectrum after the critical time $x^-_0$
and are responsible for the information transfer.}}
\end{center}

Thus a reasonable estimate of the critical line behind which
virtual boundary effects can be expected to
take place is obtained by including
this negative vacuum contribution to $T_{++}$ in the classical
equation of motion (\ref{xxt}) for the boundary.
The trajectory of this line in the $x^\pm$-plane is indicated in fig. 3.
Before the shockwave
has passed, it coincides with the line at which in the semi-classical
theory the dilaton takes its critical value $\phi = \phi_{cr}$
\cite{cghs,rst}, and afterwards goes over in the apparent horizon
of the black hole formed by the incoming matter wave \cite{rst}.
We wish to emphasize, however, that this critical line should {\it not}
be confused with the {\it location} of the reflecting boundary, as the region
behind it is still present in our model (see fig 2.). It only borders the
region of space-time from which the quantum gravitational corrections
appear to originate to an asymptotic observer.

The corrections to the Hawking spectrum will thus become visible
after the time when the out-state starts to
depend on the physics that takes place in the small region
between the critical line and the event horizon $x^- = 0$.\footnote{Note
that the distance between the critical line and the event horizon $x^- = 0$
is $1/E$ in `Planck units'.}
{}From the above discussion we see that the reflection
time $x^-_0$ on ${\cal I}_R^+$ at which this first happens is related
to the incoming time $x^+_0$ via $x^-_0 \simeq - {1/ (\lambda^2 x_0^+)}$.
This time must be compared with the initial time at which the Hawking
radiation starts to come out, which is around $x^- \simeq -\PP_+/\lambda^2$.
Thus, provided the incoming energy $E = \PP_+ x^+_0$ is much larger
than 1 (measured in units of $\lambda$), there is a semi-classical
regime in our model where the black hole will emit thermal
radiation.

The energy carried out during this time interval proportional to $\log (E)$,
and so the corrections to the Hawking spectrum will already start to
occur when the black hole has lost only a small fraction of its total mass.
This results clearly contradicts the usual supposition that strong coupling
physics starts playing a role only after the black hole has reached a mass
of the order of the Planck mass. As we will now argue, this supposition
may indeed be false in general. Namely, it can be seen
that, due to the exponentially growing redshift, any question about the
form of the asymptotic state on ${\cal I}_R^+$ at late times translates
back into a question about the form of the state near the horizon at
extremely short distances. It is indeed true that, and
Hawking calculation is based on this fact, the event horizon looks like
a perfectly regular part of space-time to a local inertial observer, so
that the state there looks approximately like the vacuum.
This changes, however, when this inertial observer wants to make
observations concerning the structure of this state at very short
distances. In that case he will clearly notice the presence of the
infalling matter, and this short distance structure will thus depend
on strong coupling physics.
So the fact that we find corrections already at early out-going
times does not contradict any (well-established)
semi-classical result. Instead it shows explicitly that to determine
what happens after a critical time on ${\cal I}^+_R$ one must go beyond the
semi-classical approximation, as from that point on Planck scale physics
will play a role. Our dynamical boundary provides a
simple model for these strong coupling effects.

\newsection{Algebraic Properties of the $S$-Matrix}

It is useful to make a comparison between the present
formulation of dilaton gravity and critical string theory.
There is evidently a large similarity between the two if we
identify the matter and $(\X^+,\X^-)$-fields with the transverse
and light-cone string coordinates, respectively. In fact, this
correspondence can be made exact when we compare our model with
critical open string theory in a constant electric field. In that
case the two light-cone string fields have an expansion similar to
(\ref{xpans}), except that the frequency sum is in general discrete
\cite{acny}. However, it can be seen that in the limit in which
the electrical field strength goes to the critical value \cite{acny}
the open string spectrum becomes effectively continuous.
Intuitively, this happens because at this value the electric field
stretches the string to cover a full quadrant of the Minkowski plane.
This correspondence with open strings in an electric field can thus
be used to introduce an infrared cut-off in our model, namely by
taking the electric field just below the critical value. After this,
standard techniques of string theory become available in studying
the properties of the Hilbert space.

In this section we will use this finite volume regularization
to make a number of precise statements about the
dilaton gravity scattering matrix defined in section 3.3.\footnote{For
recent work closely related to this section, see \cite{kazama}.}
In particular, we prove that the in- and out-modes $\alpha_i(\omega)$
and $\beta_j(\omega)$ each provide a complete basis for the physical
Hilbert space and establish the existence of a unitary
transformation between these two bases. The physical $S$-matrix is obtained
from this transformation by performing a suitable
projection within the zero-mode sector. We also show how the
computation of the $S$-matrix elements can be streamlined by using the
Virasoro decomposition of the space of physical states. As an
example, we explicitly compute the amplitude for a process that
involves particle production in the strong coupling region.

\newsubsection{Dilaton Gravity in a Finite Volume.}

Following the above discussion, we now put the fields $f_i(u)$ and
$\X^\pm(u)$ in a finite volume by identifying $u$ modulo $2\pi L$.
We use the following mode expansions
\ba
\label{expa}
\del_u f_i(u) &=& \sum_{m=-\infty}^{\infty} f_m^i
   e^{-i\frac{m}{L}u} \ ,
\nonumber\\[2mm]
\del_u \X^\pm(u) &=& \pm \sum_{m=-\infty}^{\infty} x_m^\pm
   e^{-i\frac{(m\pm i\mu)}{L}u} \ ,
\ea
where we introduced the dimensionless parameter $\mu=\lambda L$.
The canonical commutation relations of the fields (\ref{expa})
lead to
\be
\label{modecom}
[f^i_m,f^j_n] = m \,\delta_{m+n} \delta^{ij} \ , \quad
[x^+_m,x^-_n] = (m+i\mu) \,\delta_{m+n} \ .
\ee
The Virasoro generators $L_m$ for $m\neq 0$ take the usual
form; for $m=0$ there is a non-standard shift, which we can
fix by defining $L_0 = \half [ L_1, L_{-1}]$. The result is
\be
L_m = \half \sum_{i,n} :\! f^i_n f^i_{m-n} \! :
      + \sum_n :\! x^+_n x^-_{m-n} \! :
      + \half \mu^2 \delta_{m,0} \ ,
\ee
where the normal ordering signs imply that we take the
symmetric product of the zero-modes in $L_0$.
In string theory the parameter $\mu$ measures the strength of the
electric field, and the limit of critical field strength corresponds
to the limit $\mu \ra \infty$. If at the same time one takes $n \ra
\infty$ with $\omega = n/\mu$ finite one recovers the infinite volume
theory.

In order to describe the Hilbert space we shall pick
{\it in}\ and {\it out}\ vacuum states that are annihilated by all
positive frequency modes $f^i_m$, $x_m^\pm$, $m>0$.
They are then also annihilated by the all $L_n$ with $n>0$.
We will assume that there is no momentum in the matter sector,
so that the modes $f^i_0$ annihilate these vacua.
The remaining zero modes $x^+_0$ and $x^-_0$ are canonically
conjugate and need to be treated with some care. To avoid
a cluttering of indices lateron, we will write
\be
x^+_0 \equiv x \ ; \quad x^-_0 \equiv y \ .
\ee
The algebra is thus
\be
\label{zma}
[x,y] = i\mu \ .
\ee
We will look for {\it in}\ and {\it out}\ vacuum states that are
eigenstates of the operator $(xy+yx)$. The corresponding eigenvalue
can be obtained from the condition that the vacua are annihilated by
the total $L_0$-operator, which is a sum of a matter-, ghost- and
gravitational part. Just as in string theory, the ghost- and matter
vacuum have $L_0$-eigenvalue $-1$ and $0$, hence for the gravitational
$L_0$-operator we must have
\be
L_0 |0\rangle = |0\rangle.
\ee
Working out the oscillator algebra gives the following condition
for the zero modes
\be
\label{zmev}
(xy+yx) |0\rangle = (-\mu^2 + 2) |0\rangle \ .
\ee
In the coordinate representation for $x$, where we write
$y= -i\mu \frac{d}{dx}$, we can represent the vacuum
by the following wave function
\be
\psi_0(x) = x^{-i\delta -\half}\ , \qquad {\rm with} \quad
\delta=\half\mu - \frac{1}{\mu} \ .
\ee
This wave function is well-defined for $x>0$. For the out-going
states, where we use the coordinate $y$, we can similarly use
the wavefunction $\psi_0(y)=(-y)^{i\delta-\hf}$, which is
well-defined for $y<0$.

Although at the classical level the linear dilaton vacuum is
completely covered by the coordinate ranges $\X^+ >0$, $\X^-<0$, it
will turn out to be inconsistent to restrict the range of the zero modes
$x$ and $y$ the corresponding values $x>0$ and $y<0$. This means
that we need to worry about the fact that, due to the branch cut at
$x=0$, the wavefunction $\psi_0(x)$ is not uniquely defined away from
the positive real axis. We should either give a prescription for how to
continue to negative values of $x$, or more generally, introduce two
independent in-vacuum states $|0, in\rangle_{\pm}$, and similarly two
out-vacuum states $|0, out\rangle_{\pm}$, via
\ba
\label{vacua}
&& \langle x |0,in \rangle_\pm =
\frac{1}{\sqrt{2 \pi}} \, |x|^{-i\delta-\hf}\, \theta(\pm x) \ ,
\nonumber\\[2mm]
&& \langle y |0,out \rangle_\pm =
\frac{1}{\sqrt{2 \pi}} \, |y|^{i\delta-\hf}\, \theta(\pm y) \ .
\ea
These {\it in}\ and {\it out}\ vacuum states are linearly dependent,
and the change of basis from the one to the other will lead to factors in
the $S$-matrix, which we will describe later. We will also postpone
the discussion of the physical significance of this vacuum-doubling.

\newsubsection{Completeness of the Lightcone Bases}

The following states form a basis for the Fock space in the
{\it in}\ notation
\be
\label{fock}
|\{\lambda\},in\rangle_\pm =
\prod_{i,m>0} (f^i_{-m})^{\lambda_m^i}
\prod_{m>0}  (x^-_{-m})^{\lambda_m^-}
\prod_{m>0}  (x^+_{-m})^{\lambda_m^+}
\, x^{\beta-i\omega}\, |0,in\rangle_\pm \ .
\ee
Similarly, the states $|\{\lambda\},out\rangle_{\pm}$
form a basis in the {\it out}\ notation.

As in string theory, we define physical states to be those
states that are annihilated by the positive frequency
Virasoro generators $L_m$, $m>0$, and that are eigenstates
of $L_0$ with eigenvalue 1. The latter condition fixes
the value of the zero-mode exponents $\beta$, $\omega$
in (\ref{fock}) according to
\ba
\label{expts}
\beta({\lambda_m^i,\lambda_m^+,\lambda_m^-}) &=&
-\sum_m \lambda_m^+ + \sum_m \lambda_m^- \ ,
\nonumber\\[2mm]
\omega({\lambda_m^i,\lambda_m^+,\lambda_m^-}) &=&
\frac{1}{\mu}\sum_m  m(\lambda_m^i+\lambda_m^+ +\lambda_m^-)\ .
\ea
In this section we will only be concerned with states that
satisfy these conditions or the corresponding conditions
for the states in the {\it out}\ basis.

We discussed before that the physical in-states for the 24
scalar fields are created by the dressed oscillators
$\alpha^i(\omega)$. In a finite volume we write these as
$A^i_n$, defined as
\be
\label{Aos}
A_n^i = \mbox{\small{$ \frac{1}{2\pi L} $}}
        \int_0^{2\pi  L} du\, \del_u f_i(u)
        (\lambda \X^+)^{i n/\mu}\ .
\ee
Note that the integrand in here is periodic of period $2 \pi L$.
The physical out-states are created by the oscillators
\be
\label{Bos}
B_n^i = \mbox{\small{$ \frac{1}{2\pi L} $}}
        \int_0^{2\pi L} du\, \del_u f_i(u)
        (\lambda \X^-)^{-i n/\mu}
\ee
In this section we will show that every in-state, created
by repeated application of the $A$-oscillators, can be decomposed
as a linear combination of out-states written in terms
the $B$-oscillators. This decomposition is modulo spurious
physical states, \ie\ physical states that are orthogonal to
every physical state. This result implies that the $S$-matrix
elements defined as $\langle out | in \rangle$ define a unitary
$S$-matrix, provided we include the Hilbert space sectors
corresponding to both vacua.

The line of reasoning that we will follow for establishing the
completeness of the $B$-basis of physical modes is closely analogous
to Brower's proof of the no-ghost theorem for critical strings
in $D=26$ dimensions \cite{brower}. The plan of action is as
follows. Working in the out-language, where mode expansions are
defined using powers of the field $\X^-(u)$, we will first define
operators $\Xt^+_n$ and $\Phi^-_n$ which together with the dressed
oscillators $B_n^i$ span the complete Hilbert space of states with
$L_0=1$. This means that a general state can be written as a linear
combination of states $|\{\mu\},out\rangle_{\pm}$ of the form
\be
\label{mustate}
|\{\mu\},out\rangle_{\pm} =
\prod_{i,n>0} (B^i_{-n})^{\mu_n^i}
\prod_{n>0}   (\Xt^+_{-n})^{\mu_n^+}
\prod_{n>0}   (\Phi^-_{-n})^{\mu_n^-} |0,out\rangle_{\pm}\ .
\ee
We then go on to show that a state of the form (\ref{mustate}) is
physical if and only if $\mu^-_n=0$ for all $n>0$, \ie\ if there are
no factors $\Phi^-_{-n}$. In addition we find that such a physical
state is spurious as soon as it contains one or more of the operators
$\Xt^+_n$. This will establish that every physical state can be
written as a state created by using the oscillators $B^i_n$ only,
denoted by $|B,out\rangle_{\pm}$, plus a state that is
spurious and physical,
\be
\label{noghost}
|\,physical\rangle = |B,out\rangle_+ + |B,out\rangle_-
       + |\,spurious\;\; physical\rangle .
\ee
Putting a physical in-state, created by the $A_n^i$, on the left
hand side of this relation will then establish the desired result.

Before we come to the operators $\Xt^+_n$ we define operators $\Xh_n^+$ by
\ba
\Xh^+_n &=& \mbox{\small{$\frac{-\mu^2}{2\pi L}$}}
  \int_0^{2\pi L} du\;
  :\!\del \X^+ (\lambda \X^-)^{-\frac{in}{\mu}+1}\! :
  - \mbox{\small{$\frac{\mu^2}{2\pi L} (1-\frac{in}{\mu})$}}
  \int_0^{2\pi L} du\, (\lambda \X^-)^{-\frac{in}{\mu}}
\nonumber\\[3mm]
      && - \mbox{\small{$\frac{\mu}{4\pi} (1-\frac{in}{\mu})$}}
         \int_0^{2\pi L} du\, \del[\log(\del \X^-)]
         (\lambda \X^-)^{\frac{-in}{\mu}}\ .
\ea
This expression has been chosen such
that it commutes with the Virasoro generators $L_m$,
\be
[ L_m, \Xh^+_n ] = 0 \ .
\ee
This guarantees that the action of $\Xh^+_n$ on a
physical state gives another physical state. A remarkable
property is that the modes $\Xh^+_n$ define a Virasoro algebra
among themselves
\be
[ \Xh^+_m, \Xh^+_n ] = (m-n) \Xh^+_{m+n}
                   + 2 m(m^2+\mu^2)\delta_{m+n}\ .
\ee
We now define the operators $\Xt^+_n$ and $\Phi^-_n$ as
\ba
\Xt^+_n &=& \Xh^+_n + (\mu^2+1)\delta_n
          - \half \sum_{i,m} : B_{n-m}^i B_m^i :
\nonumber\\[3mm]
\Phi^-_n &=& \mbox{\small{$\frac{1}{2 \pi L}$}} \int_0^{2\pi L} du\,
             (\lambda \X^-)^{-\frac{ni}{\mu}}\ .
\ea
These operators commute with the oscillators $B_m^i$
and satisfy the following commutation relations
\ba
\label{commu}
&& [ \Xt^+_m, \Xt^+_n ] = (m-n)\, \Xt^+_{m+n} \ ,
\nonumber\\[2mm]
&& [ \Xt^+_m, \Phi^-_n ] = -n\, \Phi^-_{m+n}\ , \quad
   [\Phi^-_m, \Phi^-_n]=0 \ .
\ea
Note that the vanishing of the central term in the Virasoro
algebra satisfied by the $\Xt^+_n$ is a consequence of our
choice of taking $N=24$ scalar fields for the matter system.

We will now argue that all states of the form (\ref{mustate})
are linearly independent. We start by picking a set of oscillators
$B_n^i$ and applying them to the vacuum to create the state
$|\{\mu^i_n\},out\rangle_{\pm}$. (In the corresponding analysis of
light cone gauge string theory such states are usually called DDF
states.) As a first result, we claim all the states that we can
create from $|\{\mu^i_n\},out\rangle_{\pm}$ by acting with the
operators $\Xt^+_n$ and $\Phi^-_n$, $n<0$, are linearly independent.
At a given level $N$ in the module, where
$N=\sum n(\mu^-_n+\mu^+_n)$, this follows from the fact
that the determinant of the matrix of inner products of all
states at this level is non-vanishing. This can be established
by using only the commutation relations (\ref{commu}), the fact
that the positive modes of $\Xt^+$ and $\Phi^-$ annihilate the
state $|\{\mu^i_n\},out\rangle_{\pm}$ and the fact that $\Phi^-_0$ is
non-vanishing on $|\{\mu^i_n\},out\rangle_{\pm}$.
An elegant derivation of this result was presented by Thorn
in \cite{thorn} (see also \cite{GSW}).
A second result, which again follows simply from the
algebraic properties of the generators $B^i_n$, $\Xh^+_n$ and
$\Phi_n^-$, is that all DDF states are mutually orthogonal and
that states created by acting with $\Xt^+_n$ and $\Phi^-_n$ on
different DDF states are also mutually orthogonal. Combining these
observations shows that all the states (\ref{mustate}) are
linearly independent. An easy counting argument then shows that
all {\it out}\ states $|\{\lambda\},out\rangle_{\pm}$ with $L_0=1$
(compare with (\ref{fock}), (\ref{expts})) can be written as
linear combinations of the states $|\{\mu\},out\rangle_{\pm}$.

The fact that the commutator $[L_m, \Phi^-_{-m}]$ is non-vanishing
for $m>0$ directly implies that a state of the form (\ref{mustate})
is not physical as soon as one of the $\mu^+_n$ is nonzero. This
shows that the dressed oscillators $B^i_n$ together with the
$\Xt^+_n$, which all commute with the $L_m$, create all the physical
states in the Hilbert space. The fact that the $\Xt^+_n$ form a
centerless Virasoro algebra and that the $\Xt^+_n$ with $n\geq0$
annihilate the DDF states implies that every physical state
(\ref{mustate}) containing one or more of the $\Xt^+_n$ is
a spurious state. This finally brings us to the relation
(\ref{noghost}), which guarantees the existence of a unitary
transformation between the physical in-states and out-states.

\newsubsection{The Zero Mode Part of the $S$-matrix}

The calculation of the $S$-matrix elements is in principle
straightforward as far as the oscillator modes $f^i_n$ and
$x^\pm_n$, $n\neq 0$, are concerned. By using the expansion
\be
\Bigl(X^\pm(u)\Bigr)^{i\omega}=e^{-i\omega u}\sum_n
{1 \over n!}{\Gamma(i\omega+1)\over \Gamma(i\omega-n+1)}
(x^\pm_0)^{i\omega-n}
\Bigl(\sum_m x^\pm_m e^{-i{m\over L}u}\Bigr)^n
\ee
one can perform the $u$-integrals in (\ref{Aos}) and (\ref{Bos})
and explicitly write the physical in- and out-states in terms
of integer powers of the oscillator modes. The change of basis
from in-states to out-states (as in (\ref{noghost})) can then be
done by evaluating the appropriate inner products.\footnote{Note that,
although this calculation looks similar to that of open string amplitudes, our
$S$-matrix does {\it not} coincide with the open string $S$-matrix.}
For this one
uses the commutator algebra (\ref{modecom}) and the fact that the
vacuum is annihilated by all negative frequency modes. Working
out the algebra, one eventually is left with an expression that
only involves the zero-modes. We shall first explain how to compute
the relevant zero mode expectation values and after that discuss
the structure of more general $S$-matrix elements.

To evaluate the expressions for the $S$-matrix elements
we need to know the expectation values
\be
\label{expect}
{}_\pm\langle 0,out |\; |y|^{-i\omega^\prime-n} |x|^{-i\omega-m}\;
|0,in \rangle_{\pm}
\ee
where $\omega$ and $\omega^\prime$ are real, and $n$ and $m$ are
integers (in the finite volume $\omega$ and $\omega^\prime$ will be
of the form $N/\mu$, where $N$ is a positive integer).
It easily seen that the expectation values (\ref{expect}) are only
non-zero when $\omega=\omega^\prime$ and $m=n$. This follows from the
property (\ref{zmev}) of the physical vacuum states,
which is sufficient to fix the expectation value of integer
powers of $x$ and $y$.

Let us first put $m=n=0$ and evaluate the matrix expectation value
(\ref{expect}). The transition from $x$ to $y$ is just a Fourier
transformation, and it is easily checked that this leads to the
following relation
\be
\label{expec}
{}_b\langle 0,out |\, |y|^{-i\omega^\prime} |x|^{-i\omega}\,
|0,in \rangle_a =
S_a^b(\omega)
\ee
where $a,b=\pm$ and the coefficients are given by
\ba
\label{somega}
S_{\pm}^{\mp}(\omega) &=&
\frac{1}{\sqrt{2\pi}}\,\mu^{-i(\omega+\delta)}\,
e^{\pm\frac{\pi i}{4}} e^{-\frac{\pi}{2}(\omega+\delta)}\,
\Gamma(\hf-i(\omega+\delta))\ ,
\nonumber\\[2mm]
S^{\pm}_{\pm}(\omega) &=&
\frac{1}{\sqrt{2\pi}}\,\mu^{-i(\omega+\delta)}\,
e^{\pm\frac{\pi i}{4}} e^{\frac{\pi}{2}(\omega+\delta)}\,
\Gamma(\hf-i(\omega+\delta))\ .
\ea
The relation
\be
|S^\pm_-(\omega)|^2 + |S^\pm_+(\omega)|^2 =1
\ee
shows that the transition from {\it in}\ to {\it out}\ vacuum states
preserves probability.

By repeatedly using the zero-mode algebra (\ref{zma}) we can
finally derive
\be
{}_b\langle 0,out| \; |y|^{-i\omega^\prime-n}
|x|^{-i\omega-m} \; | 0,in \rangle_a =
(i\mu)^{-m}\,{\Gamma({\mbox{\small{$1\over 2$}}}-m-i(\omega+\delta))
\over\Gamma({\mbox{\small{$1\over 2$}}}-i(\omega+\delta))}
\,S_a^b(\omega)\, \delta_{\omega,\omega^\prime}\,\delta_{m,n} \ .
\ee

It turns out that not all these zero mode overlaps are relevant for
computing the $S$-matrix elements between asymptotic in- and out-states
of the infinite volume theory.
For observers who do not enter the strong coupling region
of space-time, but instead perform their measurements in the
asymptotic regions ${\cal I}_R^-$ and ${\cal I}_R^+$, only part of
the full zero mode wavefunction is accessible (cf. the discussion
at the end of section 2). Therefore, to describe experiments
that involve measurements in those two regions, we are forced to
project the above zero mode $S$-matrix onto the component $S_+^-$.
Due to this projection, which can be thought of as a single
$Z_2$-measurement of the sign of $x^-$,
the physical observable part of the full
dilaton gravity $S$-matrix will no longer be represented by a
unitary operator. However, it is clear from the
results of this section that no information will be lost.
We will comment further on this point in section 6.
In the following subsection we will continue work with the full
unitary zero mode $S$-matrix (\ref{somega}).

\newsubsection{Virasoro Decomposition.}

We are now ready to discuss the general structure of the
$S$-matrix elements between physical in- and out-states. We first
remark that the
space of physical in-states, which are created by the dressed
oscillators $A_m^i$, can be organized in terms of representations
of the $c=24$ Virasoro algebra generated by
\be
L^A_m = \half \sum_{i,n} : \! A_n^i A_{m-n}^i \! : \ .
\ee
Every physical in-state can be written as a linear combination of
states of the form
\be
\label{hwm}
     (L^A_{-1})^{l_1} (L^A_{-2})^{l_2} \ldots (L^A_{-M})^{l_M}
     |\, M, \Delta;in \rangle_{\pm} \ ,
\ee
where $|\, M, \Delta ;in \rangle_{\pm}$ is a highest weight state
(primary state) of the Virasoro algebra. The highest weight
state carries energy $\Delta$ and it has $SO(24)$ quantum numbers
that are determined by the structure of the tensor $M$.

In the previous section we defined the operators $\Xh^+_m$ and $\Xt^+_m$,
and we established that the latter create spurious physical states
in the Hilbert space. We can similarly define operators $\Xh^-_m$ and
$\Xt^-_m$ in the in-sector and show that modulo spurious physical states
we can write
\be
\label{t=t}
L_m^A = \Xh^-_n + (\mu^2+1)\delta_n \ .
\ee
This relation is nothing else than the quantum implementation
of the statement that the combined energy momentum tensor of the
matter fields $f^i$ and the dilaton gravity fields $\X^\pm$
should vanish. The relation (\ref{t=t}) implies that, modulo
spurious physical states, the descendant states in each highest
weight module can be created by acting with the $\Xh^-_n$, \ie,
with operators that are written entirely in terms of the dynamical
dilaton gravity fields $\X^+$ and $\X^-$. This observation can
be used to simplify the computation of the $S$-matrix elements.

We now consider the scattering matrix for an incoming state that
is primary under the $L^A_m$. We write it as
\be
\label{Astate}
|\,M,\Delta; in\rangle_{\pm} =
\sum_{I_{(1)},\ldots I_{(k)}} M_{I_{(1)},\ldots I_{(k)}}
         (A_{-1})^{I_{(1)}} (A_{-2})^{I_{(2)}} \ldots
         (A_{-k})^{I_{(k)}} |0,in\rangle_{\pm}\ ,
\ee
where $I_{(j)}$ are multi-indices: $I_{(j)}=\{ i^1_j,
i^2_j,\ldots,i^{\mu_j}_j\}$ and we write
$A_{-j}^{I_{(j)}}=A_{-j}^{i^1_j} A_{-j}^{i^2_j}, \ldots,
A_{-j}^{i^{\mu_j}_j}$. The conformal dimension of the state
is expressed as $\Delta=\sum l \mu_l$. Explicitly working out the
expressions for the $A^i_m$ oscillators for a general incoming
state leads to a complicated sum of terms, each of which is written
as a product over negative modes of the (bare) $f^i$ and $x^+$
oscillators and an appropriate power of $x=x^+_0$. However, if the
incoming state is primary the expression simplifies and reduces to
a form similar to (\ref{Astate}), with the $A^i_m$ replaced by
$f^i_m$ and an overall factor $x^{-i\omega}$, with
$\omega = \frac{\Delta}{\mu}$. (The fact that all other terms
drop out can for example be seen by working out the condition that
the state is physical; the sum of the leading terms that we just
described is physical by itself and their are no candidates for
sub-leading corrections involving only $f^i_m$ and $x^+_m$.)
Writing the same state in the out-basis of course gives a similar
result, this time with an overall factor $y^{i\omega}$.
It will be clear that the $S$-matrix on these states is simply
diagonal, the only non-trivial effect being the
change of basis in the zero-mode sector described
above. In formula
\be
|\, \Delta,M; in \rangle_a =
S_a^b\left( \mbox{\small{$\frac{\Delta}{\mu}$}} \right)
|\, \Delta,M; out \rangle_b
\ee
with the factor $S_a^b(\omega)$ as given in (\ref{somega}).
Obviously, the scattering phase only depends on the conformal
weight (energy) $\Delta$ of the incoming state and not on any
other details of the group theoretical factor $M$.

The situation becomes more interesting if we consider an incoming
state that is a descendant as in (\ref{hwm}). We will find that
in that case the scattering matrix shows a mixing of different
states at a certain level of
a given highest weight module. We will first illustrate this
phenomenon with a few examples and later discuss the more general
structure.

The simplest example of a descendant state is a state of the
form $L^A_{-1} |\, \Delta ;in \rangle$. At this level the module
has a single state, which means that the scattering will again
be diagonal. However, the $S$-matrix elements are not simply given
by the zero mode overlaps $S_a^b((\Delta+1)/\mu)$, but instead take
the form
\be
L^A_{-1} |\, \Delta; in \rangle_a =
   S_a^b \left( \mbox{\small{$\frac{(\Delta+1)}{\mu}$}}\right)
   \frac{(\mu-i)(\mu^2+i\mu+2\Delta)}{(\mu+i)(\mu^2-i\mu+2\Delta)}
   L^B_{-1} |\, \Delta; out \rangle_b \ ,
\ee
as can be checked by direct computation. For $\Delta=0$ this
expression is somewhat meaningless, since in that case the incoming and
outgoing states are spurious physical states.

At level 2 the general highest weight module has two independent
descendant states, generated by $(L_{-1}^A)^2$ and $L^A_{-2}$,
respectively. These provide the first example of matter states
that show non-trivial mixing of in the process of scattering off
the strong coupling region of the dilaton gravity system. We have
worked out the corresponding $S$-matrix elements for these particular
descendants in the module with primary state $A_{-1}^i \vac$
with $\Delta=1$. The following states provide an orthonormal basis
for the in-states
\ba
&& (e_A)_a = \frac{1}{\sqrt{13}}\left( L^A_{-2} - \half (L_{-1}^A)^2 \right)
          A_{-1}^i |0,in\rangle_a
       = \frac{1}{2 \sqrt{13}} \sum_j (A_{-1}^j A_{-1}^j )
         A_{-1}^i  |0,in\rangle_a \ ,
\nonumber\\[3mm]
&& (f_A)_a = \frac{1}{\sqrt{12}} (L_{-1}^A)^2 A_{-1}^i |0,in\rangle_a =
         \frac{1}{\sqrt{3}} A_{-3}^i |0,in\rangle_a \ .
\ea
The explicit change of basis from the incoming states
$(e_A,f_A)_a$ to the outgoing states $(e_B,f_B)_b$
yields the following $S$-matrix at this level
\be
\left( \begin{array}{c} e_A \\ f_A \end{array} \right)_a =
S_a^b\left( \mbox{\small{$\frac{3}{\mu}$}} \right)
\left( \begin{array}{cc} \lambda & \nu \\
                         \nu     & - \lambda^* \frac{\nu}{\nu^*}
       \end{array} \right)
\left( \begin{array}{c} e_B \\ f_B \end{array} \right)_b \ ,
\ee
where
\ba
\lambda &=& \frac{\mu^5-2i\mu^4+13\mu^3-32i\mu^2-4\mu-20i}
                 {(\mu+2i)(\mu^2-3i\mu+4)(\mu^2-i\mu+4)} \ ,
\nonumber\\[3mm]
\nu &=& -2i \sqrt{39}\, \frac{(\mu-i)(\mu^2+2)}
                 {(\mu+i)(\mu+2i)(\mu^2-3i\mu+4)(\mu^2-i\mu+4)} \ .
\ea
It is easily checked that this level two $S$-matrix
describes a unitary transformation. The physical $S$-matrix elements
between asymptotic in- and out-states are obtained by projecting
onto the right zero mode sectors. The off-diagonal amplitudes
proportional to $\nu$
describe the process where a single particle of discretized momentum
3 approaches the strong coupling region of the dilaton gravity
system and gets scattered into 3 out-going particles, each
of momentum 1.

With this Virasoro structure at hand, we can try to sharpen a bit the
physical picture that we developed in earlier sections.
If we substitute for a moment the term `$SO(24)$
quantum numbers' for the term `information', we see
immediately that our $S$-matrix preserves information.
This is because, as we saw, the $S$-matrix respects the
decomposition into Virasoro modules, which each carry a
specific representation of $SO(24)$. An in-going Virasoro
primary state is simply reflected to an out-going state
with identical structure, picking up an energy dependent
phase factor. However, descendant states do mix when scattering
off the strong coupling region of the dilaton gravity system.

The in-state created by $\widehat{A}_i(p_+)$ (see (\ref{ABhat})),
is a linear combination, as in (\ref{Alpha}), of states created
by the $\alpha_i(\omega)$. The latter correspond to the
finite volume states created by the $A_{-n}^i$ with $n$ large
(of order $L$), which can be written as $(L^A_{-1})^{n-1}
A_{-1}^i |0,in\rangle_a$. Since these states are `deep'
descendants, they will scatter into out-states which contain many,
less energetic, particles. This is in accordance with the result of
section 4.3, that a considerable fraction of the out-going state
describes thermal radiation. It further suggest that the information
will typically come out in the form of low energy modes.

\newsection{Discussion}

In this paper we have presented an exact quantization procedure for
two-dimensional dilaton-gravity, and we used it to construct
a scattering matrix between asymptotic in- and out-states. We have further
given strong evidence that this $S$-matrix indeed describes the
formation and evaporation of two-dimensional black holes.
In particular, we have made contact with the semi-classical theory
and the standard derivation of Hawking radiation.

Our results also give an indication when important quantum corrections
to the Hawking spectrum can be expected to appear.
Somewhat surprisingly we find that this already occurs after a relatively
short time $t= \log(M/\lambda)$ after the black hole, with mass $M$,
has been formed.
As we explained, by this time signals become visible to the asymptotic
observer that originated from sub-Planckian distances from the event
horizon. To
understand the structure of the asymptotic outgoing state it is then no
longer reliable to assume that the state near the horizon
is given by the vacuum. The reason is that the infalling matter
distorts this local vacuum state in a slight way via quantum gravitational
interactions, and although this effect is hardly noticeable to a local
inertial observer, it has important consequences for an asymptotic
observer. Since time translations at infinity correspond to boosts
near the horizon, the asymptotic state will depend on the ultra-local
fine-structure of the state in that region. In a way,
the asymptotic observer is looking at the strong coupling
physics near the horizon through a Planckian magnifying glass.
Applying the same argumentation to the 3+1 dimensional situation leads to
the prediction that already after a time
$$
t=2M \log\Bigl({M \over M_{pl}}\Bigr)
$$
quantum gravitational effects can start to modify the thermal spectrum of
Hawking radiation.
For a macroscopic black hole this is well before the time that its
mass is reduced to $M_{pl}$. Thus the instant at which quantum gravitational
corrections start to play a role depends not only on the mass of the black
hole at a given time, but also on its history.

Our dilaton gravity $S$-matrix contains the one-particle
$S$-matrix discussed in section 2 as an overall factor acting in the
space of zero modes, and as a consequence it requires that we perform a
projection within this zero mode sector to ensure that the final state
can be properly interpreted as an asymptotic state in the right out-region.
Although we have shown that this projection does not lead to any
information loss, the resulting $S$-matrix does not define a
unitarity operator in the usual sense. It is an important question to what
extent the need for this projection is an artifact of the present model, or
whether it is a general feature of all quantum gravitational models containing
black holes. It seems to indicate that the dilaton gravity model has a
quantum instability against decaying into a state that is unobservable
from outside the black hole, and that the only stabilizing mechanism is
to project, as if one is performing a measurement, on the observable
part of the wave function.  While the model is
not complete without a better understanding of this issue, it seems to
us that this apparent problem is relatively mild compared to some of the
problems that arise in other proposed ways of dealing with quantum black
holes. However, a real comparison can only be made by trying to see if
each of the other possible scenarios can be realized in the form of
a fully quantum mechanical treatment of dilaton gravity.

\medskip

\medskip

\noindent
{\bf Acknowledgements}

\noindent
We thank T. Banks, C. Callan, G. 't Hooft, I. Kogan, M. Porrati,
L. Susskind, L. Thorlacius, F. Wilczek  and E. Witten for helpful discussions.
The research of K.S. is supported by DOE-Grant DE-AC02-76ER-03072,
that of H.V. by NSF Grant PHY90-21984
and that of E.V. by an Alfred P. Sloan Fellowship, the W.M. Keck
Foundation, the Ambrose Monell Foundation, and NSF Grant PHY 91-06210.

\noindent
{\renewcommand{\Large}{\normalsize}

\end{document}